# DISCRETE THERMODYNAMICS AND EMERGENCE OF BIFURCATIONS IN CHEMICAL SYSTEMS

B. Zilbergleyt

System Dynamics Research Foundation, Chicago, USA

sdrf@ameritech.net

Bifurcations in the chemical systems originate from a tension, induced by the external thermodynamic force (TdF). This tension forces the system to move to its tolerance threshold, the bifurcation point, where the thermodynamic branch and the whole set of the system states split by two. But what about the bifurcation itself: does it snap to its final, well known form of a pitchfork, as it is usually introduced in all sources known to the author (their amount is innumerous, e.g. [1,2]), or does it emerge somewhere and then evolve? This paper intends to answer that question.

## THE STRONG AND THE WEAK CHEMICAL SYSTEMS

The Le Chatelier Response (LCR) defines a relationship between the system shift $\delta$ from "true" thermodynamic equilibrium (TdE) and external thermodynamic force causing the shift. It was introduced in discrete chemical thermodynamics [3] as

(1) $\qquad\qquad\qquad\qquad\qquad\qquad\qquad\qquad \rho_j = \Sigma \omega_p \delta_j^p$,

with the system index j and a finite set p={0,1,2, ..p}, whose length (i.e. the value of p) loosely depends upon the system complexity. It was found later, that, with regard to their behavior under the external impact, the chemical systems split into two groups, the weak and the strong, depending on the starting value – corresponding to the 0 or 1 - of the p set. This dissimilarity leads to slightly different logistic maps [4,5] which define the chemical systems states, namely

(2) $\qquad\qquad\qquad\qquad \ln[\Pi_j(\eta_j,0)/\Pi_j(\eta_j,\delta_j)] - \tau_j(w_0 - w_p \delta_j^p) = 0$,

for $p_0=0$, and

(3) $\qquad\qquad\qquad\qquad \ln[\Pi_j(\eta_j,0)/\Pi_j(\eta_j,\delta_j)] - \tau_j \delta_j (w_0 - w_p \delta_j^p) = 0$

for $p_0=1$; the weights $w_0$ and $w_p$, related to different orders of the system interaction, are unknown *a priori* and supposed to fall within the [0,1] interval. A small difference in the second term leads to a quite distinctive type of behavior. A particular case of a system with a simple transformation A*=A (the laser light emitting reaction [6])

(4) $\qquad\qquad\qquad\qquad\qquad\qquad\qquad\qquad A^*=A + h\nu$,

is exemplified in Fig.1 by graphical solutions to the maps (2) and (3) (in the form of Static Inverse Bifurcation Diagrams). The strong system shows essential resistance to the changes caused by the external force – it has the TdE area, laying on the abscissa at $\delta=0$, featuring a totally classical behavior up to the TdE limit. Then the system deviates from TdE in the area of open equilibrium (OpE) until the shift again achieves a critical value and the thermodynamic branch splits in two; here, at the bifurcation point starts the bifurcations area (the "strong system", Fig.1). The "weak system" has no TdE area and departs from "true" thermodynamic equilibrium at the smallest value of the external force, pretty soon reaching out to the bifurcation point.

In the strong system new branches diverge lazily, while in the weak system the external force is pushing them to the limit values of 0 and 1 almost vertically, right away from bifurcation point (the "weak system", Fig.1). Another definition we use for the weak system is a 'triggering system". Obviously, the strong systems are capable of a higher resistance to the external impact than the weak systems.

"Thermodynamic strength" of chemical reaction is represented by $\eta$ that depends upon its $\Delta G^0$ and the initial amounts of its participants. Influence of $\eta$ (and p as well) upon the inverse bifurcation diagrams was investigated earlier on [3,4]. The structural strength, introduced in this paper and illustrated by Fig.1, is a system characteristic because it makes sense only if the system interacts with its environment and depends upon the structure of the LCR. Presented in this paper data is to exemplify the impact of the $w_0$ and $w_1$ parameters on bifurcations in the chemical systems.

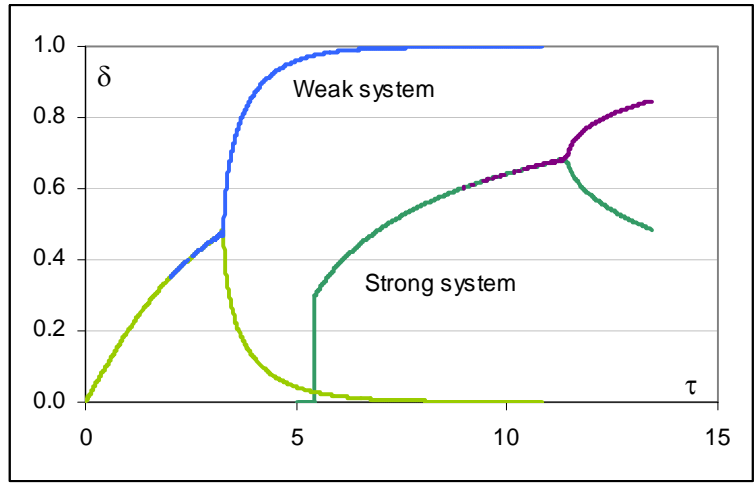

Fig.1. Graphical solutions to maps (2) (weak system, p={0,1}) and (3) (strong system, p={1}), full bifurcation diagrams period 2, both feature $\eta=0.8$, $w_0=1$, $w_1=1$.

**THE BIRTH AND DEVELOPMENT OF THE PITCHFORK BIFURCATIONS**
In this work results were obtained for reaction (4) which may be considered as merely a phase transition with release of energy. Chemical systems show similar behavior regardless their complexity. The birth and evolution of the inverse bifurcations in the strong system ($\eta=0.8$, $p=1$) with varying $w_0$ and $w_1=1$ are shown in Fig.2. In the weak system with the same parameters the picture looks pretty much the same. Fig.3 and Fig.4 show the birth and evolution of bifurcations in respectively the strong and the weak systems ($\eta=0.8$, $p=1$) with $w_0=1$ and varying $w_1$. Their similarity is remarkable – bifurcation starts at a certain value of $w_1$ as a bud on the thermodynamic branch at the endpoint of the OpE area. It slowly opens up like a contour of a flower whose sidelines then evolve gradually into traditional pitchfork bifurcation branches.

The following figures then represent the shift $\delta$ vs. the growth parameter $\tau$ (in this particular case coinciding numerically with the TdF) for the 2-level laser as a special case of the weak system, with parameters $\eta=0.999$, $p=\{0,1\}$: Fig.5 is relevant to varying $w_0$ and $w_1=1$, while Fig.6 is relevant to $w_0=1$ and varying $w_1$; there is a kind of "unrest" or a bifurcation precursor at $\delta\approx0.6$ on the upper left side curve (Fig.6). The bifurcation emergence and development very much repeat the previous pictures. Both are featuring the line spectra of transitions between the branches, with discovered in [6] gaps, which might be of a potential use to keep the 2-level laser in control. Actually, even if the system's LCR is restricted by $p=1$, the birth, the evolution and then stabilization of new branches occur in a many (at least 4) parametric space of $\delta$, $\eta$, and the LCR factors $w_0$ and $w_1$.

**CONCLUSION**
The illustrative data of this paper were intended to follow the emergence of pitchfork bifurcations and highlight the key point of their development. So far we are not aware of the precise physical meaning of the LCR statistical weights, however ability to control their values could lead to some interesting opportunities. First, a restriction on $w_0$ can allow us to avoid bifurcations in the system of interest at all or hold them close to their degeneration state (Fig.2). Similarly, as one can see in

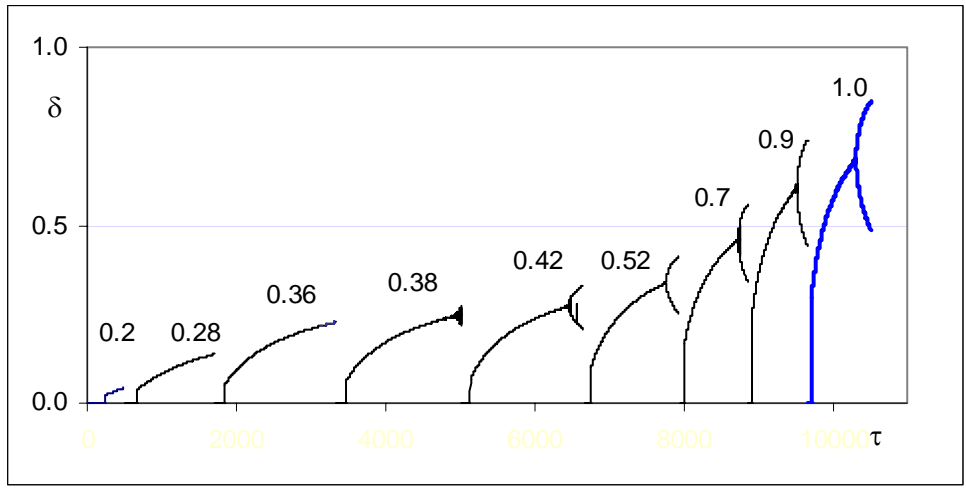

Fig.2. Strong system, part 1: varying $w_0$ (values shown at the curves), $w_1=1$.

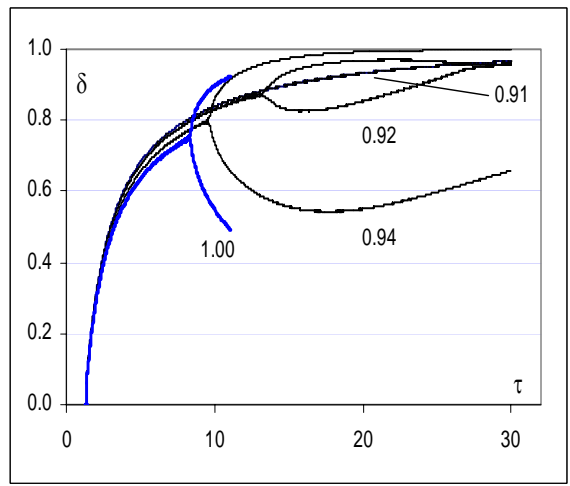

Fig.3. Strong system, part 2: $w_0=1$, varying $w_1$ (values shown at the curves).

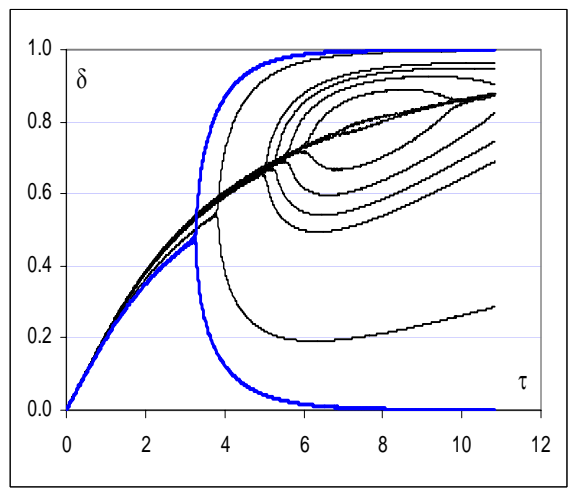

Fig.4. Weak system, part 2: $w_0=1$, varying $w_1$, internal to external curves $w_1$ values: 0.75, 0.76, 0.77, 0.78, 0.79, 0.80, 0.9, 1.0.

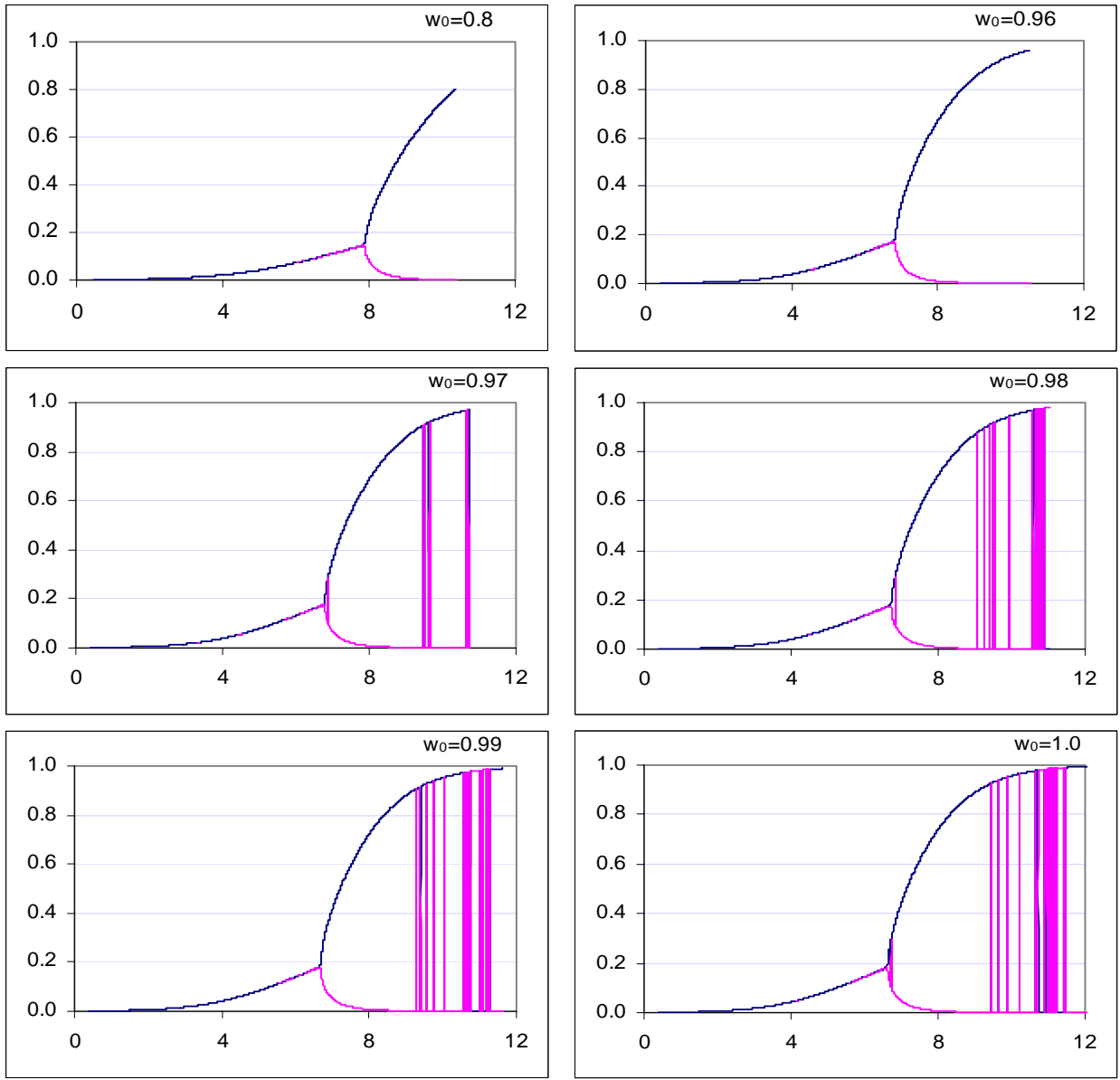

Fig.5. 2-level laser, part 1: $\delta$ vs. $\tau$, varying $w_0$ (values are shown on the boxes), $w_1=1$.

Fig.5, in the application to the 2-level laser a restriction on $w_0$ may provide us with an opportunity to turn off the lasing ability. Also, it follows from Fig.6, that control over $w_1$ provides for an opportunity to select a more convenient structure of the transition spectra to control emission, as well as to turn on and off the lasing ability (graphs for $w_1=0.46$ and $w_1=0.5$) by passing back and forth the whole bubble bifurcation. Though all this sounds very unusual, the presented data are based on a solid theory and reliable iterative simulations.

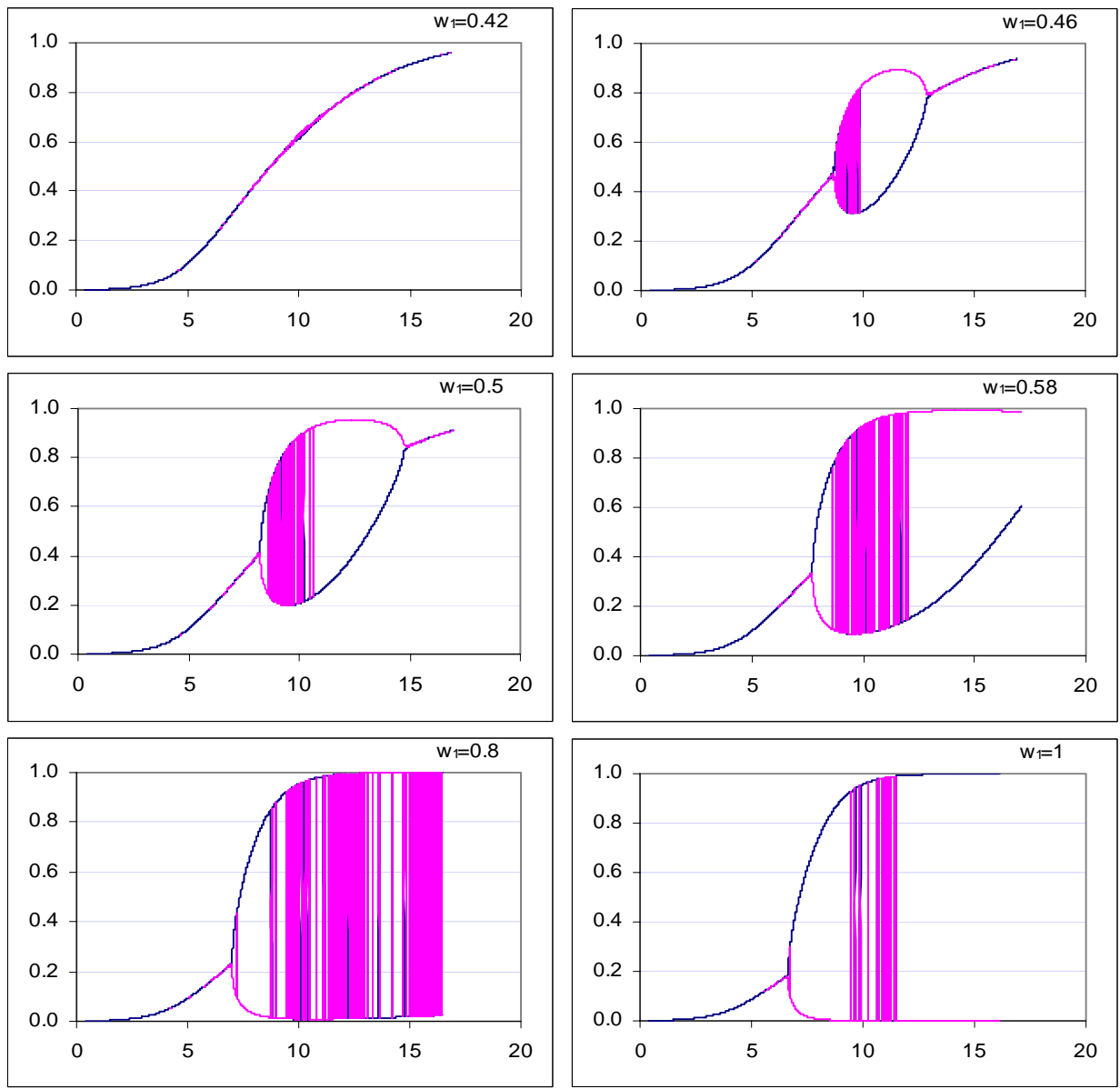

Fig.6. 2-level laser, part 2: δ vs. τ, $w_0=1$, varying $w_1$ (values are shown on the boxes).

**REFERENCES**
1. I. Epstein, J. Pojman. An Introduction to Nonlinear Chemical Dynamics. Oxford University Press, New York, 1998.
2. C. Beck, F. Schlögl. Thermodynamics of Chaotic Systems. Cambridge University Press, Cambridge, UK, 1993.
3. B. Zilbergleyt. Domain of States of Chemical Systems: Le Chatelier Response, Structure of the Domains and Evolution. arXiv.org:physics /0513133.
4. B. Zilbergleyt. Chemical Equilibrium as Balance of Thermodynamic Forces. arXiv.org:physics/0404082.
5. B. Zilbergleyt. Discrete Thermodynamics of Chemical Equilibria. 19$^{th}$ ICCT, Boulder, 2006.
6. B. Zilbergleyt. Discrete Thermodynamics of 2-level laser - Why *Not* and When *Yes*. arXiv.org:physics /0609044.